\documentclass[conference]{IEEEtran}

\usepackage{amsfonts}
\usepackage[cmex10]{amsmath}
\usepackage{graphicx}
\usepackage{setspace} 
\usepackage{times} 
\usepackage{epsfig}
\usepackage{subfigure} 
\usepackage{latexsym} 
\usepackage{cite}
\usepackage{epsfig}
\usepackage{fixltx2e}
\usepackage{verbatim}
\usepackage{makeidx}
\usepackage{cases}
\newtheorem{theorem}{Theorem}
\newtheorem{proposition}{Proposition}
\newtheorem{conj}[theorem]{Conjecture}
\newcommand{\defeq}{\stackrel{\mbox{{\tiny def}}}{=}}
\def\Var{{\rm Var}\,}
\def\E{{\rm E}\,}

\ifCLASSINFOpdf
\else
\fi 
\begin{document}

\title{Exploiting Self-Interference Suppression for Improved Spectrum Awareness/Efficiency in Cognitive Radio Systems}

\author{\IEEEauthorblockN{Wessam Afifi and Marwan Krunz}

\IEEEauthorblockA{Department of Electrical and Computer Engineering, University of Arizona\\
E-mail: \{wessamafifi, krunz\}@email.arizona.edu}}

\maketitle
\begin{abstract}
Inspired by recent developments in full-duplex communications, we propose and study new modes of operation for cognitive radios with the goal of achieving improved primary user (PU) detection and/or secondary user (SU) throughput. Specifically, we consider an opportunistic PU/SU setting in which the SU is equipped with partial/complete self-interference suppression (SIS), enabling it to transmit and receive/sense at the same time. Following a brief sensing period, the SU can operate in either simultaneous transmit-and-sense (TS) mode or simultaneous transmit-and-receive (TR) mode. We analytically study the performance metrics for the two modes, namely the detection and false-alarm probabilities, the PU outage probability, and the SU throughput. From this analysis, we evaluate the sensing-throughput tradeoff for both modes. Our objective is to find the optimal sensing and transmission durations for the SU that maximize its throughput subject to a given outage probability. We also explore the spectrum awareness/efficiency tradeoff that arises from the two modes by determining an efficient adaptive strategy for the SU link. This strategy has a threshold structure, which depends on the PU traffic load. Our study considers both perfect and imperfect sensing as well as perfect/imperfect SIS. 

\end{abstract}
\IEEEpeerreviewmaketitle
\section{Introduction}

Until recently, the idea that a wireless device can transmit and receive simultaneously on the same frequency channel, i.e., operate in full-duplex (FD) mode, was deemed impossible. The traditional scenario was that at a given time, a node can transmit or receive, but not both, which is often called half-duplex (HD) operation. The problem of achieving FD communications is that the transmitted power from a given node is typically much larger than the received power of another signal at the same node. While the node is receiving, its transmitted signal is considered as self-interference.
\let\thefootnote\relax\footnote{This research was supported in part by NSF (under grants CNS-1016943 and CNS-0904681, IIP-0832238, IIP-1231043), Raytheon, and the "Connection One" center. Any opinions, findings, conclusions, or recommendations
expressed in this paper are those of the authors and do not necessarily reflect the views of the National Science Foundation.}

The infeasibility of FD communications have recently been challenged in several efforts, which have successfully demonstrated the possibility of FD communications using self-interference suppression (SIS) techniques \cite{choi_achieving_single_channel,duarte_fullduplex_wireless_communications,bozidar_rethinking_indoor}. The main task in these works is to suppress self-interference to a level that enables FD communications. Recent studies \cite{jain_practical_realtime_fullduplex,sahai_pushing_limits} have shown that a transmitting device can significantly suppress its own interference by up to 80 dB, enabling it in certain scenarios to concurrently transmit and receive. 

There are two main approaches for SIS: RF interference cancellation and digital baseband interference cancellation. Combined, these two approaches may still not achieve the amount of SIS required for FD communication. In \cite{choi_achieving_single_channel}, the authors proposed an antenna-based SIS technique and used it in conjunction with the two previous techniques to reach the required suppression limit. In this technique, two appropriately spaced transmit antennas and one receive antenna are used to ensure that the transmitted signals add destructively at the receive antenna and cancel each other. This technique has some limitations in terms of design complexity (number and placement of antennas) and the destructive interference points that will appear in the far field. Furthermore, there is a bandwidth constraint and a practical limitation on the operation of such a scheme, as it requires manual tuning. The authors in \cite{jain_practical_realtime_fullduplex} addressed these limitations and proposed an interference cancellation mechanism based on signal inversion. This technique has some practical limitations too, as discussed in \cite{jain_practical_realtime_fullduplex}. Another technique for SIS was presented in \cite{duarte_fullduplex_wireless_communications}, where the authors explored antenna placement as an additional cancellation technique to analog and digital interference cancellation. Some aspects of designing the physical and MAC layers with SIS are discussed in \cite{singh_efficient_fair_mac,sahai_pushing_limits}. 

While advances in SIS are being sought aggressively, exploiting FD/SIS in network protocol design is still in its very early stages. To support statistical quality-of-service (QoS), the authors in \cite{cheng_resource_allocation} proposed an optimal resource allocation scheme for wireless FD and HD relay networks. They showed that the optimal capacity with FD is not always twice that of the HD mode, and that a hybrid transmission mode may achieve better performance than using FD alone. Cross-layer optimization for routing in FD-capable wireless networks was studied in \cite{fang_routing_FD}. The authors considered the problem of selecting end-to-end routes, first to maximize the total profit of users subject to node constraints, and secondly to minimize the power consumption subject to rate demands.

Opportunistic spectrum access (OSA) is one of the prevalent means for improving spectrum efficiency \cite{FCC_2002}. In OSA, secondary users (SUs) sense the spectrum and opportunistically access it if the primary users (PUs) are thought to be idle \cite{Zhao_07_survey,haykin}. In \cite{cheng_FD_sensing}, a continuous-time Markov chain model was used to analyze the achievable throughput of PUs and SUs under an FD spectrum sensing scheme. Using this scheme, the authors showed that PUs can maintain their required throughput and SUs can increase their achievable throughput compared with the throughput under the HD scheme. The authors in \cite{cheng_imperfect_fullduplex} derived the false-alarm and detection probabilities for an FD spectrum sensing scheme, assuming a non-slotted cognitive radio (CR) network. Some factors such as signal bandwidth, antennas placement error, and the amplitude difference of the transmit signals were analyzed. It was shown that the unavoidable error due to signal bandwidth has little impact on network performance. Hence, the FD scheme can be effectively used in CRs.

In this paper, we consider a CR setting in which the receiver of an SU uses SIS techniques to mitigate the undesirable interference from its own transmitter. This SIS capability can be utilized in several ways. It can be used to increase the SU throughput by enabling bidirectional simultaneous {\em transmission-reception (TR)}. It can also be used to increase the SU's awareness of PU activity by allowing the SU to sense while transmitting, which we refer to as the {\em transmission-sensing (TS)} mode. We investigate the efficient policy for an SU link, taking into consideration the tradeoff between spectrum efficiency (throughput) and spectrum awareness (detection capability). Our objective is to determine the optimal action for an SU link, whose aim is to maximize its throughput subject to a given PU outage probability. We also attempt to find the optimal sensing and transmission durations for this SU link. An important step towards reaching this goal is to design and formulate appropriate performance metrics for the SU network.

The contributions of this paper are as follows. First, we propose two novel modes of operation for opportunistic SUs with SIS capability: TS and TR. Second, we derive the detection and false-alarm probabilities, the PU/SU collision probability, and the SU throughput for both modes. Based on these metrics, we compare the performance of the two modes with the traditional HD transmission-only (TO) mode. Third, we study the sensing-throughput tradeoff for CRs in both TS and TR modes. Specifically, for both modes we determine the ``optimal" sensing and transmission durations that maximize the SU throughput subject to constraints on the PU outage probability. Fourth, we explore the spectrum awareness/efficiency tradeoff that arises due to the competing goals of minimizing the collision probability with the PU (TS mode) and maximizing the SU throughput (TR mode). Our objective here is to determine an efficient strategy for the SU link that enhances its throughput subject to a given collision probability. Our scheme has a threshold-based structure, which depends on the PU traffic load: For low traffic loads, the SU should operate in the TR mode, whereas the TS mode is superior at high loads. Finally, we study the impact of perfect and imperfect sensing with perfect/imperfect SIS. To the best of our knowledge, this is the first paper to address the spectrum awareness/efficiency tradeoff in CRs that arises from the new modes of operations, TS and TR.

The rest of the paper is organized as follows. The system model and operation modes are described in Section \ref{system_model}. In Section \ref{spectrum_sensing}, we derive the detection and false-alarm probabilities for the two modes, and formulate the corresponding outage probabilities. The sensing-throughput tradeoff is discussed in Section \ref{problem_formulation}. We explore the spectrum awareness/efficiency tradeoff and determine the appropriate transmission strategy for an SU link in Section \ref{problem_formulation}. Numerical results are presented in Section \ref{numerical_results}, followed by conclusions in Section \ref{conclusions}.

\section{System Model and Operation Modes}\label{system_model}
\subsection{System Model with SIS Capability}
\begin{figure}[tbp]
	\centering
		\includegraphics[width=0.48\textwidth]{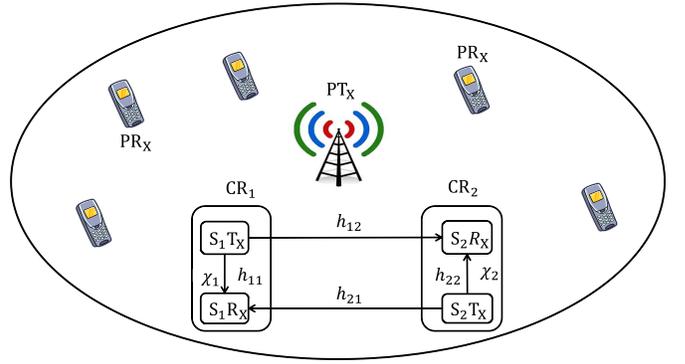}
	\caption{System model for an SU link that opportunistically accesses the spectrum of a PU network. Each SU $i$ consists of a transceiver with a given SIS capability factor $\chi_i$.}
	\label{fig:system_model}
\end{figure}

As shown in Figure \ref{fig:system_model}, we consider an SU link that opportunistically accesses a PU-licensed channel. The SU has partial/complete SIS capability, allowing it to transmit and receive/sense at the same time. Let $\chi$ be a factor that represents the degree of SIS at an SU node, $\chi \in [0,1]$. If $\chi=0$, the SU can completely suppress its self-interference; otherwise, it can only suppress a fraction $1-\chi$ of its self-interference (imperfect SIS). $\chi$ may differ from one node to another, depending on the employed SIS technique. For simplicity, we assume that $\chi$ is the same for all SUs.

For SU $i$, let $P_i$ denote its transmission power. Without loss of generality, we assume that only one SU link can be active at a given time, over a given frequency channel, and in a given neighborhood. Time/frequency scheduling for a set of SU links has been well-studied in the literature (see \cite{yuan_mobihoc_07}), and will not be addressed in this paper. For the wireless channel, we consider a path-loss model \cite{vu_scaling_laws}. The channel gain $h_{ij}$ between a transmitter $i$ and a receiver $j$ at distance $d_{ij}$ is $h_{ij}=C \, d_{ij}^{-\eta}$, where $C$ is a frequency-dependent constant and $\eta$ is the path-loss exponent. In general, $h_{ij} \neq h_{ji}$. 

A collision between PU and SU transmissions occurs whenever a secondary transmission overlaps by any period of time with a primary transmission. We assume that the PU activity (and hence, channel availability for the SU) behaves as an alternating busy/idle (ON/OFF) process. Let $Y$ be the length of the PU idle period. We assume that $Y$ is exponentially distributed with parameter $\lambda_{\mbox{\tiny\em{OFF}}}$.


\subsection{SU Operation Modes}

In addition to the classic HD mode, the SU can dynamically switch to one of two FD modes: TS and TR. The SU can decrease its collision probability or achieve higher throughput by utilizing the SIS and FD capabilities in the TS and TR modes, respectively. The description of the various modes is as follows:

\subsubsection{Transmission-Only (TO) Mode}
As shown in Figure \ref{fig:notations_1}, in this mode the SU senses the spectrum for a duration $T_{S0}$ and then carries out data transmission. The transmission duration is denoted by $T$. This is the traditional HD mode of operation, which is well studied in the literature.

\subsubsection{Transmission-Sensing (TS) Mode}
Even though FD provides the capability to transmit and sense at the same time, the SU must initially sense in a HD fashion for a duration $T_{S0}$, as shown in Figure \ref{fig:notations_2}. Based on the sensing outcome, the SU can decide whether to transmit for $T$ seconds (if the PU is idle) and simultaneously continue sensing to detect the return of a PU, or not transmit if the PU is sensed to be busy. While transmitting, the SU performs $m$ sensing actions $T_{Si}$, $i \in \left\{1,2,\ldots,m\right\}$. Thus, in the TS mode, we have $m+1$ sensing durations. If at the end of any given sensing period PU activity is detected, the SU aborts its transmission until the next cycle (which also starts with a sensing-only period of length $T_{S0}$).

\subsubsection{Transmission-Reception (TR) Mode}
\begin{figure}[tbp] 
\centering
\subfigure[Transmission-Only mode]
{
    \includegraphics[width=0.2\textwidth]{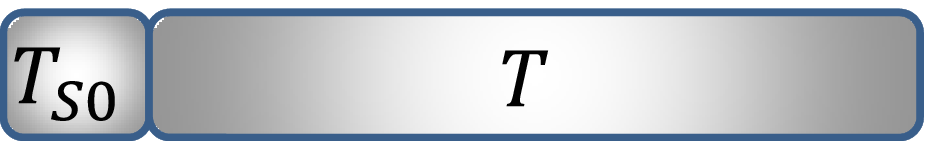}
    \label{fig:notations_1}
} \\
\subfigure[Transmission-Sensing mode]
{
		\includegraphics[width=0.2\textwidth]{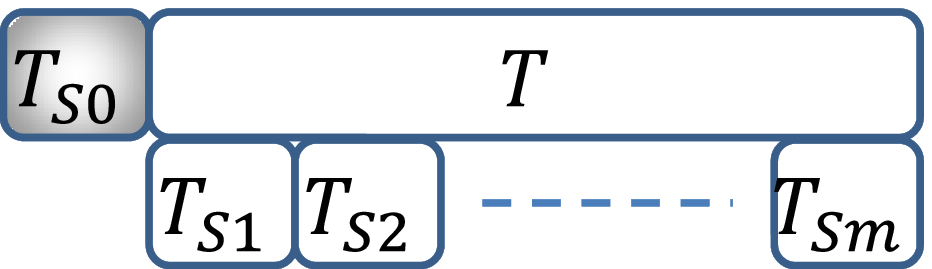}
    \label{fig:notations_2}
} \\
\subfigure[Transmission-Reception mode]
{
		\includegraphics[width=0.2\textwidth]{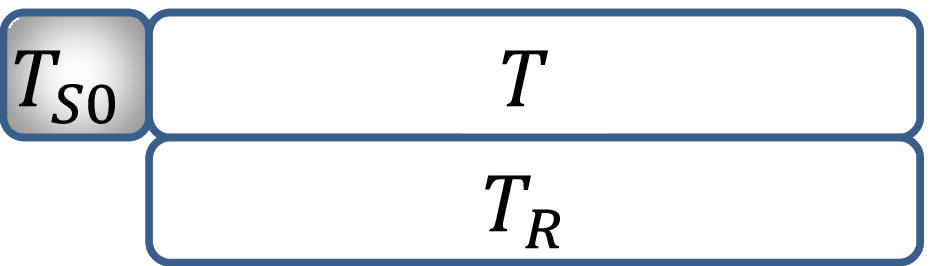}
    \label{fig:notations_3}
} \\
\caption{Different modes of operation for the SU. \subref{fig:notations_1} is the traditional HD mode, \subref{fig:notations_2} and  \subref{fig:notations_3} are considered as FD modes, although they contain a sensing-only period at the beginning.}
\label{fig:notations}
\end{figure}

Instead of sensing while transmitting, the SU may start receiving data from its peer while transmitting to that same peer, as shown in Figure \ref{fig:notations_3}. As before, an initial sensing period of length $T_{S0}$ is needed to determine channel availability. Let $T_R$ be the reception duration.

To analyze the various modes of communications, we assume that the time axis is divided into frames, where each frame consists of a sensing-only period $T_{S0}$ and a potential transmission period $T$, as shown in Figure \ref{fig:notations}.
\section{Sensing Metrics and PU Outage}\label{spectrum_sensing}
In OSA networks, SUs utilize spectrum sensing to determine the idle/busy state of a channel. SIS can be exploited to enable simultaneous transmission and sensing over the same channel (via the TS mode), which impacts the detection and false-alarm probabilities. The reason is that, in practice SIS techniques cannot completely suppress self-interference $(\chi > 0)$. Therefore, we have to account for the residual self-interference when deriving the false-alarm and detection probabilities. In this case, the hypothesis test of whether the channel is busy or not can be formulated as follows: 
\begin{subnumcases}{r(n)=} 
\chi \, s(n)  + w(n) & $H_0$ (PU idle) \label{H_0}
\\
l(n) + \chi \, s(n)  + w(n) & $H_1$ (PU busy) \label{H_1}
\end{subnumcases}
\noindent where $r(n)$ is the discretized received signal after performing spectrum sensing in the FD case, $s(n)$ is the SU's own transmitted signal before applying SIS (assumed to be a zero mean iid random signal with variance $\sigma_s^2$), $w(n)$ is the noise signal (assumed to be a zero mean Gaussian iid random process with variance $\sigma_w^2$), and $l(n)$ is the received PU signal (assumed to be a zero mean iid random process with variance $\sigma_l^2$). For simplicity, we ignore the path loss between the SU's transmitter and its reception {\em at the same node} (i.e., $h_{11}=h_{22}=1$ in Figure \ref{fig:system_model}).

The hypotheses in (\ref{H_0}) and (\ref{H_1}) are applicable to the $m$ sensing actions $T_{Si}$, $i =1,2,\ldots,m$, of the TS mode. As for the sensing-only period $\left(T_{S0}\right)$, which is used in the three modes, the hypothesis test is given by:
\begin{subnumcases}{\tilde{r}(n)=} 
w(n) & $H_0$ (PU idle) 
\\
l(n) + w(n) & $H_1$ (PU busy)
\end{subnumcases}
\noindent where $\tilde{r}(n)$ is the received signal in the HD case.

The detection probability $(P_d)$ and the false-alarm probability $(P_f)$ are defined as the probabilities that the sensing process determines the channel to be busy given $H_1$ and $H_0$, respectively. To maintain a certain level of protection for the PU, $P_d$ must be high. Increasing $P_d$ reduces the SU/PU collision probability, which has a positive effect on the PU's throughput. Hence from the PU side, the only parameter of interest is $P_d$. On the other hand, the SU should care about both $P_d$ and $P_f$. The lower the $P_f$, the higher the SU throughput, as fewer transmission opportunities will be missed. The value of $P_d$ also plays a noticeable role in determining the SU's throughput, since colliding with the PU will result in fewer successful SU transmissions. In summary, a good detection system should have a low $P_f$ and a high $P_d$.

\subsection{Energy Detection}
In the following analysis, we focus on energy-detector based sensing. The main idea is to compute the average energy of $N$ samples of the signal $r(n)$ and compare it with a threshold $\gamma$ to determine whether the PU is idle or not. The decision metric $M$ for the energy detector is defined as:
\begin{equation}
M= \frac{1}{N} \sum_{n=1}^{N} \left|r(n)\right|^2.
\end{equation}
In the FD case, $P_f$ and $P_d$ are given by:
\begin{equation}
P_f^{\mbox{\small\em({FD})}}=\Pr \left[ M > \gamma / H_0 \right]=1-F_{M/H_0}(\gamma)
\label{decision_metric_false_alarm}
\end{equation}
\begin{equation}
P_d^{\mbox{\small\em(FD)}}=\Pr \left[ M > \gamma / H_1 \right]=1-F_{M/H_1}(\gamma)
\label{decision_metric_detection}
\end{equation}
\noindent where $F_{M/H_0}(\gamma)$ and $F_{M/H_1}(\gamma)$ are the conditional CDFs of the random variable $M$ given hypothesis $H_0$ and $H_1$, respectively. 

Using the central limit theorem, we can obtain the distribution of $M$ given the two hypothesis $H_0$ and $H_1$. 

\begin{proposition}
For a large $N$, the pdf of $M$ given $H_0$ can be approximated by a Gaussian distribution with the following mean and variance:
\begin{equation}
\E[M/H_0] \defeq \mu_{M/H_0}= \chi^2 \sigma_s^2 + \sigma_w^2
\end{equation}
\begin{equation}
\begin{split}
\mbox{\em{\Var}}[M/H_0] & \defeq \sigma_{M/H_0}^2= \frac{1}{N} \Big[\chi^4 \E\left|s(n)\right|^4 + \E\left|w(n)\right|^4 \\
& \quad \quad \quad \quad \, - \left( \chi^2 \sigma_s^2-\sigma_w^2\right)^2   \Big].
\label{variance}
\end{split}
\end{equation}
\end{proposition}

To compare with the HD case \cite{sensing_throughput_tradeoff}, we assume the noise signal $w(n)$ to be circularly symmetric complex Gaussian (CSCG) and $s(n)$ a complex PSK-modulated signal. In this case, $ \E\left|w(n)\right|^4 = 2 \sigma_w^4 $ and $ \E\left|s(n)\right|^4 = \sigma_s^4 $. Substituting these values in (\ref{variance}), we get $\sigma_{M/H_0}^2= \frac{1}{N} \left(2 \chi^2 \sigma_s^2 \sigma_w^2 + \sigma_w^4\right)$. Note that the number of samples $N$ is a function of the sensing duration $T_{Si}$, $i \in \left\{1,2,\ldots,m\right\}$:
\begin{equation}
N=T_{Si} f_S
\end{equation}
where $f_S$ is the sampling rate. Accordingly, the false-alarm probability in the FD case can be expressed as:

\begin{equation}
P_f^{\mbox{\small\em(FD)}}= Q \left( \left(\frac{\gamma}{\sigma_w^2} - \chi^2 \alpha_s -1\right)\sqrt{\frac{N}{2 \chi^2 \alpha_s+1}} \right)
\label{false_alarm_FD}
\end{equation}

\noindent where $\alpha_s=\sigma_s^2/\sigma_w^2$ is the received signal-to-noise ratio (SNR) of the SU, measured at the secondary receiver of the same node, and $Q$ is the complementary distribution function of a standard Gaussian random variable.

\begin{proposition}
Under hypothesis $H_1$ and for a large $N$, the pdf of $M$ can be approximated by a Gaussian distribution with the following mean and variance:
\begin{equation}
\E[M/H_1] \defeq \mu_{M/H_1}= \sigma_l^2 + \chi^2 \sigma_s^2 + \sigma_w^2
\end{equation}
\begin{equation}
\begin{split}
\mbox{\em{\Var}}[M/H_1]& \defeq \sigma_{M/H_1}^2  = \frac{1}{N} \Big[\E\left|l(n)\right|^4 + \chi^4 \E\left|s(n)\right|^4  \\
&\!+\E \! \left|w(n)\right|^4-\left(\sigma_l^2 - \chi^2 \sigma_s^2-\sigma_w^2\right)^2 + 4 \chi^2 \sigma_s^2 \sigma_w^2 \Big].
\end{split}
\end{equation}
\end{proposition}

A similar argument to the one in \cite{sensing_throughput_tradeoff} can be used to prove the previous propositions. Suppose that $l(n)$ is also a complex PSK-modulated signal. Substituting $ \E\left|l(n)\right|^4 = \sigma_l^4 $, we get $\sigma_{M/H_1}^2= \frac{1}{N} (2 \chi^2 \sigma_s^2 \sigma_w^2 +2 \chi^2 \sigma_s^2 \sigma_l^2 +2 \sigma_l^2 \sigma_w^2 + \sigma_w^4)$. Let $\alpha_l=\sigma_l^2/\sigma_w^2$ be the received SNR of the PU, measured at the SU receiver. The detection probability in the FD case can be expressed as:

\begin{equation}
\begin{split}
P_d^{\mbox{\small\em(FD)}} &= Q \Bigg( \left(\frac{\gamma}{\sigma_w^2} - \chi^2 \alpha_s - \alpha_l   -1\right) \times \\
&\sqrt{\frac{N}{2 \chi^2 \alpha_s + 2 \chi^2 \alpha_s \alpha_l + 2 \alpha_l   +1}} \Bigg).
\label{detection_FD}
\end{split}
\end{equation}

In the HD case, where there is no self-interference, the false-alarm and detection probabilities are readily available \cite{sensing_throughput_tradeoff}:

\begin{equation}
P_f^{\mbox{\small\em(HD)}}= Q \left( \left(\frac{\gamma}{\sigma_w^2} -1\right)\sqrt{N} \right)
\label{false_alarm_HD}
\end{equation}

\begin{equation}
P_d^{\mbox{\small\em(HD)}}= Q \left( \left(\frac{\gamma}{\sigma_w^2} - \alpha_l-1\right)\sqrt{\frac{N}{2 \alpha_l+1}} \right).
\label{detection_HD}
\end{equation}

Note that under perfect SIS $(\chi=0)$, the equations for the false-alarm and detection probabilities for the FD case in (\ref{false_alarm_FD}) and (\ref{detection_FD}) converge to those of the HD case in (\ref{false_alarm_HD}) and (\ref{detection_HD}), given a specific sensing duration.

\subsection{Outage Probability}
In this section, we analyze the PU outage probability. Our analysis is conservative because it considers any time overlap between the PU and SU transmissions as outage. In this case, the outage probability is the same as the collision probability between the SU and PU transmissions. We consider the situation under perfect and imperfect sensing for each of the three communication modes. 

Generally, there are two possible events that could lead to a collision, as shown in Figure \ref{fig:collision}. First, due to its imperfect sensing, the SU may wrongly decide that the PU is idle and proceed to transmit data when the PU is actually ON.	Second, the SU may start transmitting while the PU is idle, but later on the PU switches from OFF to ON during the SU's transmission. Both events will be considered in the following analysis.

\subsubsection{TO Mode}
Under perfect sensing, collision occurs only if the PU was idle, but later on became active after the SU started its transmission (before the end of $T$). Let $\tau$ be the forward recurrence time for the PU OFF period, observed after the initial sensing-only period $T_{S0}$. $\tau$ has an exponential distribution with parameter $\lambda_{\mbox{\tiny\em{OFF}}}$ (due to the memoryless property of the exponential distribution). Hence, the collision probability in the perfect sensing case is given by:
\begin{equation}
P_{\mbox{\small\em{TO}}} = \Pr\left[\tau \leq T\right] \defeq F_{\tau}(T)= 1-e^{-\lambda_{\mbox{\tiny\em{OFF}}} T}.
\label{equation:collision_transmission_perfect}
\end{equation}

On the other hand, under imperfect sensing, the two aforementioned reasons for collision must be considered. Let $\beta$ be the PU traffic load (activity factor), $0 < \beta < 1$. The collision probability in the imperfect sensing case is given by:
\begin{equation}
\widehat{P}_{\mbox{\small\em{TO}}} =\frac{ \beta \left(1-P_d^{\mbox{\small\em(HD)}}\right)+ \left(1-\beta\right) \left(1-P_f^{\mbox{\small\em(HD)}}\right)F_{\tau}(T) } {W}
\end{equation}
\noindent where $W$ is the probability that the outcome of the initial sensing process is $H_0$, and is given by:
\begin{equation}
W = \beta \left(1-P_d^{\mbox{\small\em(HD)}}\right)+ \left(1-\beta\right) \left(1-P_f^{\mbox{\small\em(HD)}}\right).
\end{equation}
Note that $W$ is also the probability that the SU will attempt a transmission.

\subsubsection{TS Mode}
The collision probability in the perfect sensing case for this mode is simply equal to zero. The reason is that the SU is continuously sensing, and its sensing is perfect. We assume that the sensing period is small enough for a collision to occur within the sensing duration, and that the SU can quickly detect the change in the PU state \cite{Zhao_Quickest_Change_Detection,zhao_3}. 


\begin{figure}[tbp]
\centering
\subfigure[Transmission-Only (TO) mode]
{
    \includegraphics[width=0.31\textwidth]{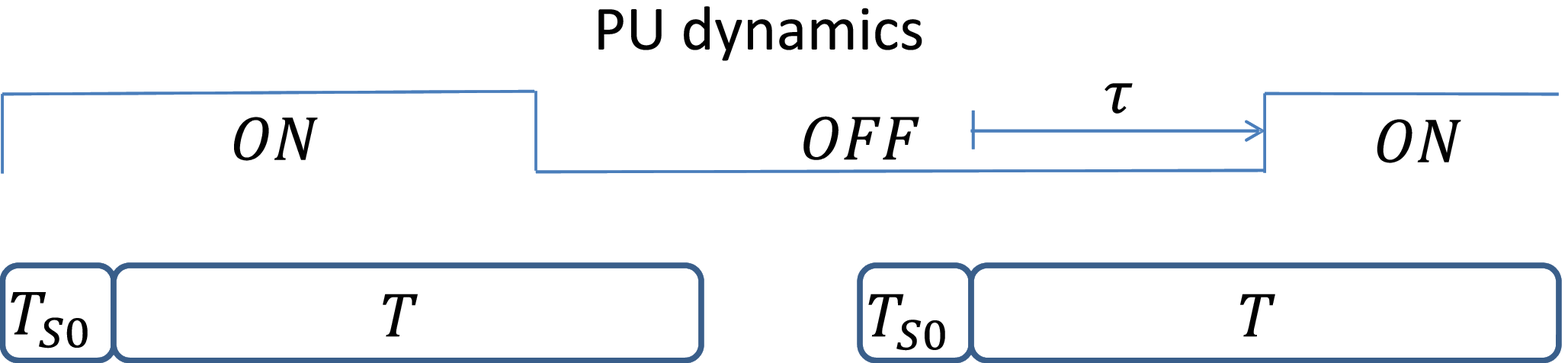}
    \label{fig:collision_1}
}
\subfigure[Transmission-Sensing (TS) mode]
{
		\includegraphics[width=0.31\textwidth]{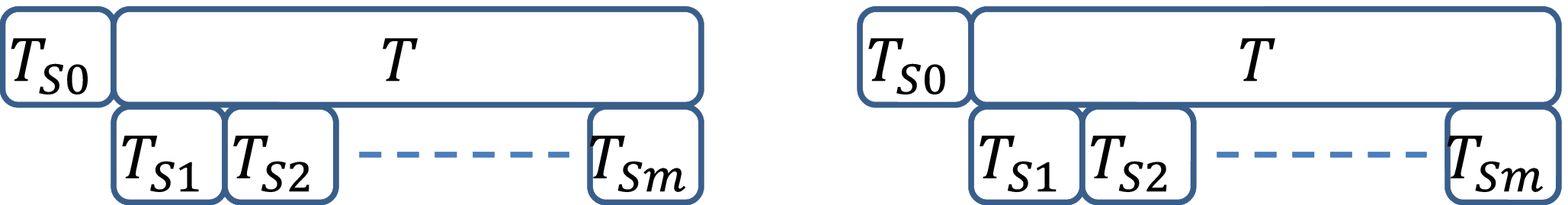}
    \label{fig:collision_2}
}
\subfigure[Transmission-Reception (TR) mode]
{
		\includegraphics[width=0.31\textwidth]{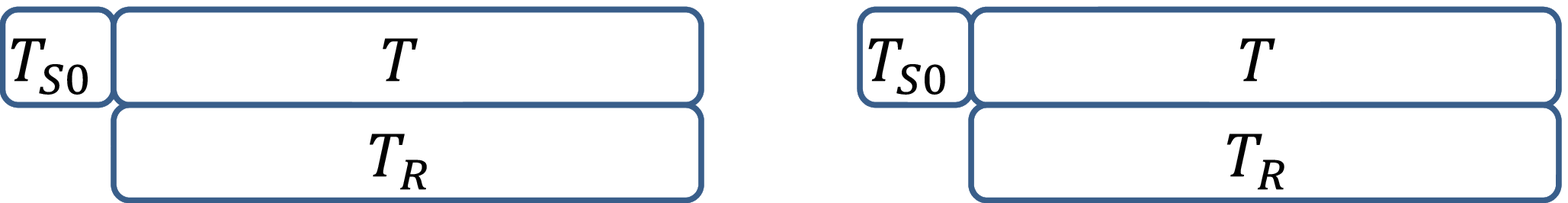}
    \label{fig:collision_3}
}
\caption{Two possibilities for collision in the three modes. $\tau$ is the forward recurrence time for the PU OFF period when observed at the end of the sensing period $T_{S0}$.}
\label{fig:collision}
\end{figure}
Consider now the imperfect sensing case. As explained in Figure \ref{fig:collision_2}, there are two scenarios for collision, which have different features than those of the TO mode. The first scenario occurs if the SU makes a wrong decision after the initial sensing period $T_{S0}$, and determines the channel to be idle when it is not. This happens with probability $A \defeq \beta \left(1-P_d^{\mbox{\small\em(HD)}}\right)$. Define $\mathbb{P}_f=\left[P_{f, 0} \,\, P_{f, 1} \,\, \ldots \,\,P_{f, m}\right]$ as an $(m+1)$ dimensional vector that represents the false-alarm probability for the $m+1$ sensing periods in the TS mode. The second scenario for collision occurs when the outcome of the initial sensing period $T_{S0}$ is $H_0$ and the PU is OFF at that time, but it later switches from OFF to ON. This may happen during any of the sensing periods $T_{Si}$, $i=1,2, \ldots ,m$. It may happen during $T_{S1}$ with probability $(1-\beta)(1-P_{f, 0}) \Pr[\tau \leq T_{S1}]$. It may happen during $T_{S2}$ with probability $(1-\beta)(1-P_{f, 0})(1-P_{f, 1})\Pr[T_{S1} \leq \tau \leq T_{S1}+T_{S2}]$, and so on. In general, we can write the $m+1$ collision possibilities as:
\begin{equation}
\begin{split}
B& \defeq (1-\beta)(1-P_{f, 0}) F_{\tau}(T_{S1}) +(1-\beta) (1-P_{f, 0}) \\
& \quad \, \, (1-P_{f, 1}) \left[ F_{\tau}(T_{S1}+T_{S2}) - F_{\tau}(T_{S1}) \right] +\ldots\\
& = (1-\beta) \sum_{i=1}^{m+1} \Bigg[ \left\{ F_{\tau} \left(\sum_{k=1}^{i}  T_{Sk}\right) - F_{\tau}\left(\sum_{l=1}^{i-1} T_{Sl}\right) \right\} \times  \\
& \quad \, \quad\quad\quad\quad\quad\quad\quad\quad\quad \prod_{j=0}^{i-1} (1-P_{f, j}) \Bigg].  
\end{split} 
\end{equation}

\noindent Note that $P_{f, 0}=P_f^{\mbox{\small\em(HD)}}$ by definition. Assuming that $T_{Si}$ is the same $\forall i \in \left\{1,2,\ldots,m\right\}$, then $P_{f, i}=P_f^{\mbox{\small\em(FD)}}$. Hence, we get:
\begin{equation}
\begin{split}
B&= (1-\beta) \left(1-P_f^{\mbox{\small\em(HD)}}\right) \sum_{i=1}^{m+1} \bigg[ \left(1-P_f^{\mbox{\small\em(FD)}}\right)^{(i-1)} \times \\
 & \quad \left\{ F_{\tau} \left(i T_{S}\right) - F_{\tau}\left((i-1) T_{S}\right) \right\} \bigg].  
\end{split} 
\end{equation}

Accordingly, the collision probability in the TS mode under imperfect sensing is given by:
\begin{equation}
\widehat{P}_{\mbox{\small\em{TS}}} = \frac{ A+B } {W}.
\label{collision_TS_imperfect}
\end{equation}

Clearly, the collision probability under imperfect sensing in the TS mode is smaller than that of the TO mode, which is the gain of using SIS in the TS mode.

\subsubsection{TR Mode}
Exploiting SIS in the TR mode primarily impacts the SU throughput, and has no effect on the collision probability. Therefore, the collision probabilities in the TR mode for perfect and imperfect sensing are similar to those of the TO mode, as shown in Figure \ref{fig:collision_3}.
\section{Adaptive SU Communication Strategy}\label{problem_formulation}

In this section, we explore how an FD-capable SU adapts its communication strategy so as to maximize its throughput without exceeding a certain outage probability. First, we study the traditional sensing-throughput tradeoff for the FD modes (TS and TR). Then, we explore a novel spectrum awareness/efficiency tradeoff that results from the TS and TR modes. Finally, we propose an efficient adaptive strategy for the SU link, which allows it to switch between the TS and TR modes.

\subsection{Sensing-Throughput Tradeoff}
First, we analyze the SU throughput under the three different modes of operation. Given our definition of a successful SU transmission (no overlap between the SU and PU transmissions), we formulate the SU throughput as the probability that no collision occurs with the PU multiplied by the maximum achievable throughput. Note that the SU may be able to communicate successfully even when the PU is ON. However, the throughput achieved without collision will dominate.
\subsubsection{TO Mode}
In the traditional HD mode, the secondary throughput under perfect and imperfect sensing can be written as:
\begin{equation}
R_{\mbox{\small\em{TO}}} = \left(1-P_{\mbox{\small\em{TO}}}\right) \frac{T}{T+T_{S0}} \log\left(1+\mbox{\em{SNR}}_{\mbox{\small\em{TO}}}\right)
\label{R_T_P}
\end{equation}
\begin{equation}
\widehat{R}_{\mbox{\small\em{TO}}} = \left(1-\widehat{P}_{\mbox{\small\em{TO}}}\right) \frac{T}{T+T_{S0}} \log\left(1+\mbox{\em{SNR}}_{\mbox{\small\em{TO}}}\right)
\label{R_TO_I}
\end{equation}
\noindent where $\mbox{\em{SNR}}_{\mbox{\small\em{TO}}}$ is the SNR at a receiving node $j$ from a transmitting node $i$. This SNR is given by:
\begin{equation}
\mbox{\em{SNR}}_{\mbox{\small\em{TO}}} = \frac {P_i \left|h_{ij}\right|^2} {\sigma_j^2}
\end{equation}
\noindent where $\sigma_j^2$ is the noise variance of node $j$.

As shown in (\ref{R_T_P}) and (\ref{R_TO_I}), the expression for the throughput is the same in the perfect and imperfect sensing cases, except for the collision probability. Henceforth, we focus on the SU throughput under imperfect sensing.

\subsubsection{TS Mode}
The formulation of the SU throughput in the TS mode is similar to the TO mode except for the collision probability. Recall that the TS mode has a lower collision probability than the TO mode. Hence, the SU throughput is given by:
\begin{equation}
\widehat{R}_{\mbox{\small\em{TS}}} = \left(1-\widehat{P}_{\mbox{\small\em{TS}}}\right) \frac{T}{T+T_{S0}} \log\left(1+\mbox{\em{SNR}}_{\mbox{\small\em{TS}}}\right)
\end{equation}
where 
\begin{equation}
\mbox{\em{SNR}}_{\mbox{\small\em{TS}}} = \mbox{\em{SNR}}_{\mbox{\small\em{TO}}}.
\end{equation}

\subsubsection{TR Mode}
The benefit of using SIS in this mode is achieving higher SU throughput, due to transmitting and receiving over the same channel. The throughput in this case will be the summation of the throughputs achieved in the two directions. It is given by:
\begin{equation}
\begin{split}
\widehat{R}_{\mbox{\small\em{TR}}} & = \left(1-\widehat{P}_{\mbox{\small\em{TR}}}\right) \Bigl[ \frac{T}{T+T_{S0}} \log\left(1+\mbox{\em{SNR}}_{\mbox{\small\em{TR}}}^{(j)}\right) \\
& + \frac{T_R}{T_R+T_{S0}} \log\left(1+\mbox{\em{SNR}}_{\mbox{\small\em{TR}}}^{(i)}\right) \Bigl]
\end{split}
\end{equation}
\noindent where $\widehat{P}_{\mbox{\small\em{TR}}}$ is the collision probability under imperfect sensing for the TR mode. The SNR in the TR mode at node $j$ is given by:
\begin{equation}
\mbox{\em{SNR}}_{\mbox{\small\em{TR}}}^{(j)} = \frac {P_i \left|h_{ij}\right|^2 } {\sigma_j^2 + \chi_j^2 P_j \left|h_{jj}\right|^2}.
\label{SNR_TR_B}
\end{equation}

Note that $h_{jj}$ is the channel gain from transmitter $j$ to receiver $j$ at the same node (i.e., the self-interference channel). Since the distance between the transmitter and receiver of the same node is quite small, path-loss is ignored in this case. That is, the only factor that affects the strength of this self-interference signal is the SIS capability factor $\chi$. 

If $T=T_R$, then
\begin{equation}
\begin{split}
\widehat{R}_{\mbox{\small\em{TR}}} & = \left(1-\widehat{P}_{\mbox{\small\em{TR}}}\right) \frac{T}{T+T_{S0}} \Bigl[ \log \left(1+\mbox{\em{SNR}}_{\mbox{\small\em{TR}}}^{(j)}\right) \\
& + \log\left(1+\mbox{\em{SNR}}_{\mbox{\small\em{TR}}}^{(i)}\right) \Bigl].
\end{split}
\end{equation}

Now that the SU throughput is obtained for each mode, we proceed to optimize the SU operation. Two optimization problems $(P1$ and $P2)$ are considered, which explore the sensing-throughput tradeoff in the TS and TR modes. We consider the imperfect sensing case in our formulation. Specifically, our objective in $P1$ is to determine the optimal sensing and transmission durations, $\mathbb{T}_S$  and $T$, so as to maximize the SU throughput in the TS mode subject to a constraint on the PU outage probability. Formally,
\begin{equation*}
\begin{aligned}
P1\!:\,& \underset{\mathbb{T}_S, T}{\text{maximize}}
& & \widehat{R}_{\mbox{\small\em{TS}}}=\left(1-\widehat{P}_{\mbox{\small\em{TS}}}\right) \frac{T}{T+T_{S0}} \log\left(1+\mbox{\em{SNR}}_{\mbox{\small\em{TS}}}\right) \\
& \text{subject to}
& & \qquad\qquad\qquad \widehat{P}_{\mbox{\small\em{TS}}} \leq \widehat{P}_{\mbox{\small\em{TS}}}^*  \qquad\qquad\qquad 
\end{aligned}
\end{equation*}
\noindent where $\mathbb{T}_S=[T_{S0} \,\, T_{S1} \,\,\ldots\,\,T_{Sm}]$ is an $(m+1)$ dimensional vector, whose elements are the sensing durations in the TS mode. $\widehat{P}_{\mbox{\small\em{TS}}}^*$ is the constraint on the outage probability.

$P1$ addresses the sensing-throughput tradeoff from different perspectives. First, for the optimization of $\mathbb{T_S}$, we have two different optimization parameters: the sensing-only period $T_{S0}$ and the $m$ sensing periods in the TS mode. For $T_{S0}$, there is an optimal solution that maximizes our objective function, because increasing $T_{S0}$ will monotonically increase the detection probability, ultimately satisfying constraint $\widehat{P}_{\mbox{\small\em{TS}}}^*$, while decreasing $T_{S0}$ will increase the transmission duration to maximize the throughput (assuming that the SU either senses or transmits over a channel). 


On the other hand, the $m$ sensing periods must only satisfy the constraint on the collision probability. In contrast to $T_{S0}$, they do not have any effect on the transmission duration because these $m$ sensing periods are done in parallel with $T$. Because sensing is conducted while transmitting, the SU will be able to achieve a lower collision probability and satisfy the constraint. Hence, increasing the transmission duration will increase the SU throughput. However, if this value is increased beyond a certain limit, it will cause a reduction in the throughput.

In the next formulation $P2$, our objective is to determine the optimal sensing and transmission/reception durations, $T_{S0}$  and $T$  respectively, to maximize the SU throughput in the TR mode subject to a given outage probability:
\begin{equation*}
\begin{aligned}
P2\!:\,& \underset{T_{S0}, T}{\text{maximize}}
& & \widehat{R}_{\mbox{\small\em{TR}}}=\left(1-\widehat{P}_{\mbox{\small\em{TR}}}\right) \frac{T}{T+T_{S0}} \times \\
& & & \Bigl[ \log\left(1+\mbox{\em{SNR}}_{\mbox{\small\em{TR}}}^{(j)}\right) + \log\left(1+\mbox{\em{SNR}}_{\mbox{\small\em{TR}}}^{(i)}\right) \Bigl] \\
& \text{subject to}
& & \qquad\qquad\qquad \widehat{P}_{\mbox{\small\em{TR}}} \leq \widehat{P}_{\mbox{\small\em{TR}}}^*  \qquad\qquad\qquad
\end{aligned}
\end{equation*}
Using a similar argument as in $P1$, it is easy to see that the sensing-throughput tradeoff exists in $P2$ w.r.t. both optimization parameters $T_{S0}$ and $T$. However, in $P2$ we only have the initial sensing duration $T_{S0}$ instead of the vector $\mathbb{T}_S$. The formulation in $P2$ is for equal transmission and reception durations. If $T \neq T_R$, it is intuitive that the solution for the optimization problem will return the same optimal value for both parameters.

Since $P1$ and $P2$ are nonconvex problems, we use a brute-force search method to find the optimal durations that maximize our objective functions. 

\subsection{Spectrum Awareness/Efficiency Tradeoff}
The TS and TR modes give rise to a spectrum awareness/efficiency tradeoff. That is, the SU may select the TS mode to continuously sense the channel of interest while transmitting to decrease the probability of collision with the PU. On the other hand, the SU may decide to utilize the spectrum efficiently by transmitting and receiving data over the same channel (TR mode). Our objective is to determine the optimal action for the SU. To do that, we consider a combined $P1-P2$ formulation, which we refer to as $P3$. In $P3$, the SU calculates the achievable throughput in the TS and TR modes under the specified constraints. It then selects the action that provides the higher throughput. The maximum achievable throughput for the SU can be stated as follows:
\begin{equation*}
 \quad\quad\quad\quad\quad\quad\quad P3\!: \widehat{R}=\text{max}\left(\widehat{R}_{\mbox{\small\em{TS}}}, \widehat{R}_{\mbox{\small\em{TR}}}\right).  \quad\quad\quad\quad\quad\quad\quad 
\end{equation*}

Let the action space of the SU be denoted by $A=\left\{a: 1(\text{TR}), 0(\text{TS})\right\}$.
\begin{conj}
The following SU strategy returns the maximum throughput:
\begin{equation}
a^*=
\begin{cases}
1 & \text{(TR)} \quad\quad\quad\quad \text{if} \, \beta < \beta^*\\
0 & \text{(TS)} \quad\quad\quad\quad  \text{otherwise}\\
\end{cases}
\end{equation}
\end{conj}
The scheme has a threshold-based structure that depends on the PU traffic load $\beta$. The SU selects the TR action if $\beta$ is smaller than a threshold value $\beta^*$, because in this case, there is a high probability that the PU will be idle and there is no need to sense the spectrum while transmitting. On the other hand, if $\beta \geq \beta^*$, the SU selects the TS mode, as the sensing process will output a `busy' outcome with high probability. Hence, the SU proceeds to sense the spectrum while transmitting, allowing it to determine the actual state of the PU and vacate the channel if the PU is sensed busy during transmission.
\section{Numerical Results}\label{numerical_results}
Unless stated otherwise, we use the following parameter values for the numerical results. We set the sampling frequency $f_S$ to 6 MHz, $\alpha_s=20$ dB, $\alpha_l=-15$ dB, $\lambda_{\mbox{\tiny\em{OFF}}}=0.01$, $\eta=4$, and $\beta=0.5$. The number of sensing periods during $T$ in the TS mode is $m=500$.

\subsection{Performance Metrics}

\subsubsection{False Alarm and Detection Probabilities}

\begin{figure}[tbp]
	\centering
		\includegraphics[width=0.37\textwidth]{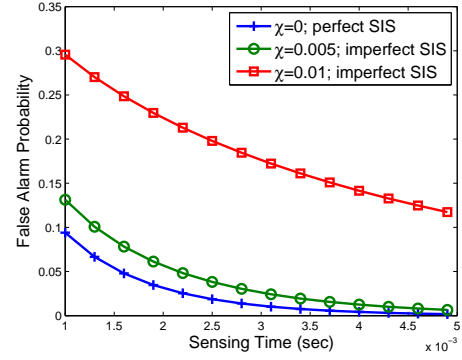}
	\caption{False alarm probability vs. sensing time in the FD case at different values of $\chi$.}
	\label{fig:false_alarm}
\end{figure}

\begin{figure}[tbp]
	\centering
		\includegraphics[width=0.37\textwidth]{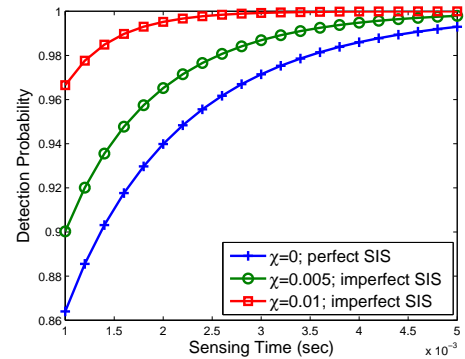}
	\caption{Detection probability vs. sensing time in the FD case at different values of $\chi$.}
	\label{fig:detection}
\end{figure}

Figures \ref{fig:false_alarm} and \ref{fig:detection} depict the false-alarm and detection probabilities in the FD case, as a function of the sensing duration at different values of the SIS capability factor $\chi$. The false-alarm probability generally decreases with the sensing duration, because a long sensing duration will result in a more reliable outcome regarding the PU state. In Figure \ref{fig:false_alarm}, the false-alarm probability with perfect SIS converges to the HD case, as expected. However, as $\chi$ increases the false-alarm probability increases (i.e., performance degrades), which is intuitive because as $\chi$ increases the interference power increases.

The detection probability generally increases with the sensing duration, because a long sensing period translates into a large number of samples, which helps in determining the actual state of the PU. As shown in Figure \ref{fig:detection}, the detection probability in the FD case also converges to that of the HD under perfect SIS. With imperfect SIS, we notice that as $\chi$ increases the detection probability increases (i.e., performance improves). The reason is that under imperfect SIS, the residual self-interference increases the average energy resulting in higher detection probability.


\begin{figure}[tbp]
	\centering
		\includegraphics[width=0.37\textwidth]{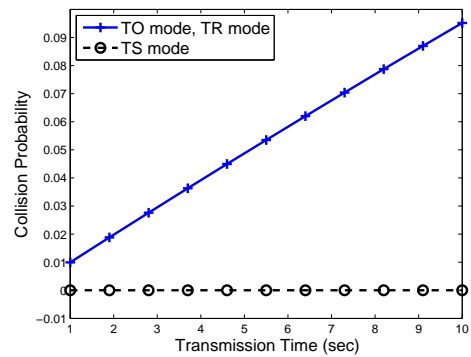}
	\caption{Collision probability vs transmission time under perfect sensing for TO, TS, and TR modes.}
	\label{fig:collision_perfect}
\end{figure}

\begin{figure}[tbp]
	\centering
		\includegraphics[width=0.4\textwidth]{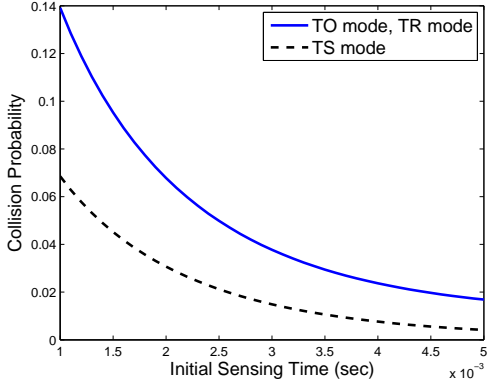}
	\caption{Collision probability vs. initial sensing time under imperfect sensing for TO, TS, and TR modes.}
	\label{fig:collision_imperfect}
\end{figure}

\begin{figure}[tbp]
	\centering
		\includegraphics[width=0.4\textwidth]{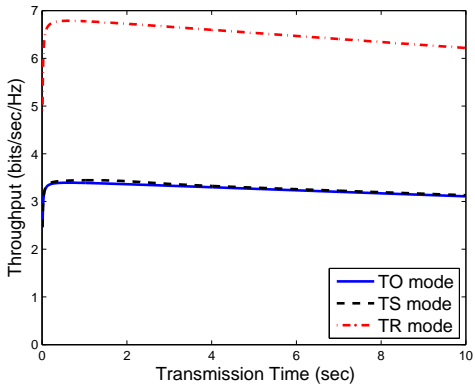}
	\caption{SU throughput vs. transmission time for imperfect sensing and perfect SIS $(T_{S0}=$ 4 msec).}
	\label{fig:throughput_imperfect_T_T_long}
\end{figure}

\subsubsection{Outage Probability}
As shown in Figures \ref{fig:collision_perfect} and \ref{fig:collision_imperfect}, the collision (outage) probabilities for the TO and TR modes are the same due to having similar sensing structures and because the reception duration in the TR mode is done in parallel with the transmission time. The SU achieves its lowest collision probability in the TS mode. The collision probability for the the TO and TR modes increases with $T$, as shown in Figure \ref{fig:collision_perfect}. The reason is that the probability that the PU becomes active again during $T$ increases with the increase in $T$, which is the only parameter affecting the collision probability in the perfect sensing case. Under imperfect sensing, increasing the sensing duration results in gaining more information about the actual state of the PU, and hence achieving a lower collision probability, as shown in Figure \ref{fig:collision_imperfect}. We also notice that the effect of varying $\chi$ on the collision probability in the TS mode is almost negligible, as the ratio of collided packets to the total transmitted packets remains almost the same, irrespective of $\chi$. The figure is omitted due to space limit.

\subsection{Sensing-Throughput Tradeoff}

\begin{figure}[tbp]
	\centering
		\includegraphics[width=0.4\textwidth]{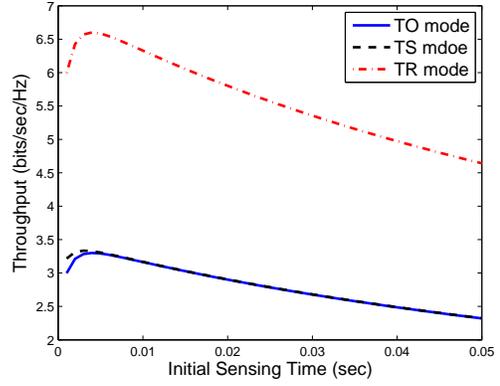}
	\caption{SU throughput vs. initial sensing time under imperfect sensing and perfect SIS $(T=$ 100 msec).}
	\label{fig:throughput_imperfect_T_S0}
\end{figure}

In Figures \ref{fig:throughput_imperfect_T_T_long} and \ref{fig:throughput_imperfect_T_S0}, we set $\mbox{\em{SNR}}_{\mbox{\small\em{TO}}}=15$ dB. It is observed that the maximum throughput is achieved in the TR mode. Notice also that the SU achieves higher throughput in the TS mode than in the TO mode due to a lower collision probability.

The sensing-throughput tradeoff is illustrated in Figure \ref{fig:throughput_imperfect_T_T_long}. We notice that increasing the transmission time $T$ initially increases the SU throughput, up to a certain point. Beyond this point, increasing $T$ increases the collision probability, which has a dominant (negative) effect on the throughput.

In Figure \ref{fig:throughput_imperfect_T_S0} we notice that increasing the sensing duration improves the SU performance by increasing the detection probability and decreasing the false-alarm probability, leading to a lower collision probability and higher throughput. However, by increasing the sensing duration, the throughput is also decreased due to the reduction in $T$ (assuming that the SU is either sensing or transmitting).

\begin{figure}[tbp]
	\centering
		\includegraphics[width=0.4\textwidth]{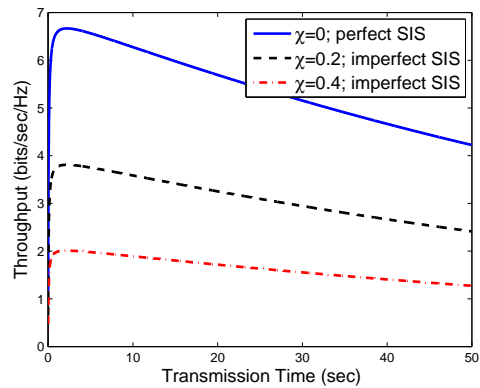}
	\caption{SU throughput vs. transmission time under imperfect sensing and perfect/imperfect SIS for the TR mode $(T_{S0}=$ 50 msec).}
	\label{fig:throughput_imperfect_imperfect_SIS}
\end{figure}

The SU throughput in the TR mode at different values of $\chi$ under imperfect sensing is shown in Figure \ref{fig:throughput_imperfect_imperfect_SIS} as a function of $T$. We notice that as $\chi$ increases, the SU throughput decreases due to the additional interference.

\subsection{Spectrum Awareness/Efficiency Tradeoff}
\begin{figure}[tbp]
	\centering
		\includegraphics[width=0.4\textwidth]{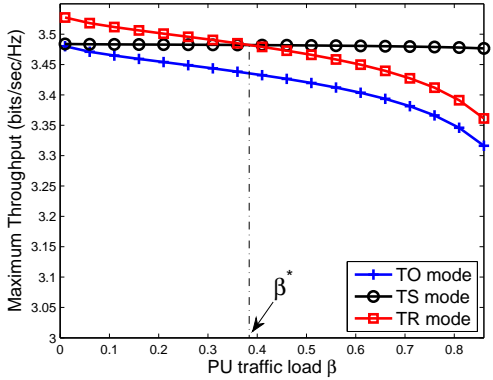}
	\caption{Maximum achievable throughput vs. PU traffic load under imperfect sensing and imperfect SIS $(\chi=0.235)$. Collision probability constraint set to 0.04.}
	\label{fig:throughput_imperfect_imperfect_SIS_traffic_load_markers}
\end{figure}

Next we consider the optimization problems $P1$ and $P2$ with a collision probability constraint 0.04 and $\chi=0.235$. Solving these problems, we found that the optimal initial sensing durations are 6.6 msec and 7 msec for the TS and TR modes, respectively. We also found the optimal transmission durations for the TS and TR modes to be 1.28 sec and 0.83 sec, respectively. We then solved $P3$. Figure \ref{fig:throughput_imperfect_imperfect_SIS_traffic_load_markers} depicts maximum achievable throughput vs. $\beta$, under imperfect SIS, where we found that $\beta^*=0.38$. If $\beta$ is high, the best action for the SU is the TS mode. On the other hand, if $\beta < \beta^*$ , it is better for the SU to transmit and receive data at the same time (i.e., operate in the TR mode) because it is highly likely that the PU will be idle.

\begin{figure}[tbp]
	\centering
		\includegraphics[width=0.4\textwidth]{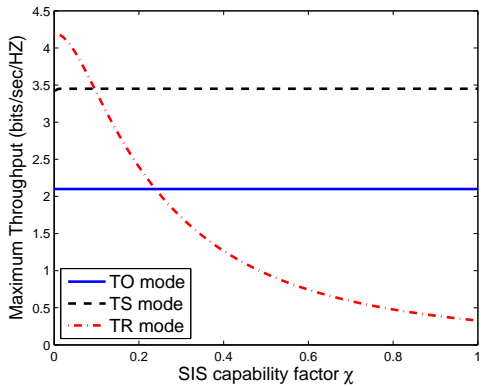}
	\caption{Maximum achievable throughput vs. $\chi$ under imperfect sensing and perfect/imperfect SIS. Collision probability constraint $=10^{-4}$ and $\beta=0.3$.}
	\label{fig:throughput_imperfect_imperfect_SIS_chi}
\end{figure}

To show the relation between the maximum achievable throughput and the SIS factor, we solve our optimization problems at different values of $\chi$ and for a collision probability constraint $=10^{-4}$. As shown in Figure \ref{fig:throughput_imperfect_imperfect_SIS_chi}, at low $\chi$, where the SU is capable of suppressing most of its self-interference, the best action for the SU is the TR mode. However as $\chi$ increases, the throughput achieved at the TR mode will decrease due to the increased self-interference. In this case, the best action for the SU will be the TS mode. 

\section{Conclusions}\label{conclusions}

In this paper, we proposed and studied a novel application of FD/SIS in the context of CRs. Two modes of operation for the SU (TS and TR) were analyzed, along with the traditional half-duplex TO mode. We found that the SU can improve its throughput and/or detection capability while operating in the TS/TR modes. We also studied the sensing-throughput tradeoff for these modes and found the optimal durations that maximize the SU throughput given a constraint on the outage probability. We explored the spectrum awareness/efficiency tradeoff and proposed an efficient adaptive strategy for the SU link. This strategy has a threshold structure that depends on the PU traffic load. We noticed that the false-alarm and detection probabilities increase with the SIS capability factor. For SUs with weak SIS capability, it is better to operate in the TS mode.

Several interesting directions for future work exist. A power control scheme is needed for the FD modes, when multiple SUs with different SIS capability factors are present. Also, the number of sensing periods within a transmission duration may be optimized to return the minimum collision probability. We will also consider the appropriate MAC design under the TS and TR modes.

\bibliographystyle{IEEEtran}
\bibliography{IEEEabrv,MyLib}

\begin{thebibliography}{10}
\providecommand{\url}[1]{#1}
\csname url@samestyle\endcsname
\providecommand{\newblock}{\relax}
\providecommand{\bibinfo}[2]{#2}
\providecommand{\BIBentrySTDinterwordspacing}{\spaceskip=0pt\relax}
\providecommand{\BIBentryALTinterwordstretchfactor}{4}
\providecommand{\BIBentryALTinterwordspacing}{\spaceskip=\fontdimen2\font plus
\BIBentryALTinterwordstretchfactor\fontdimen3\font minus
  \fontdimen4\font\relax}
\providecommand{\BIBforeignlanguage}[2]{{%
\expandafter\ifx\csname l@#1\endcsname\relax
\typeout{** WARNING: IEEEtran.bst: No hyphenation pattern has been}%
\typeout{** loaded for the language `#1'. Using the pattern for}%
\typeout{** the default language instead.}%
\else
\language=\csname l@#1\endcsname
\fi
#2}}
\providecommand{\BIBdecl}{\relax}
\BIBdecl

\bibitem{choi_achieving_single_channel}
J.~I. Choi, M.~Jain, K.~Srinivasan, P.~Levis, and S.~Katti, ``Achieving single
  channel, full duplex wireless communication,'' in \emph{Proc. of the {ACM
  Mobicom'10} Conf.}, Sep. 2010, pp. 1--12.

\bibitem{duarte_fullduplex_wireless_communications}
M.~Duarte and A.~Sabharwal, ``Full-duplex wireless communications using
  off-the-shelf radios: {F}easibility and first results,'' in \emph{Proc. of
  the {ASILOMAR'10} Conf.}, Nov. 2010, pp. 1558--1562.

\bibitem{bozidar_rethinking_indoor}
B.~Radunovic, D.~Gunawardena, P.~Key, A.~Proutiere, N.~Singh, V.~Balan, and
  G.~Dejean, ``Rethinking indoor wireless mesh design: {L}ow power, low
  frequency, full-duplex,'' in \emph{Fifth IEEE Workshop on Wireless Mesh
  Networks}, 2010.

\bibitem{jain_practical_realtime_fullduplex}
M.~Jain, J.~I. Choi, T.~Kim, D.~Bharadia, S.~Seth, K.~Srinivasan, P.~Levis,
  S.~Katti, and P.~Sinha, ``Practical, real-time, full duplex wireless,'' in
  \emph{Proc. of the {ACM Mobicom'11} Conf.}, Sep. 2011, pp. 301--312.

\bibitem{sahai_pushing_limits}
A.~Sahai, G.~Patel, and A.~Sabharwal, ``Pushing the limits of full-duplex:
  Design and real-time implementation,'' \emph{Technical Report TREE1104, Feb.
  2011, Rice University}.

\bibitem{singh_efficient_fair_mac}
N.~Singh, D.~Gunawardena, A.~Proutiere, B.~Radunovic, H.~Balan, and P.~Key,
  ``Efficient and fair mac for wireless networks with self-interference
  cancellation,'' in \emph{Proc. of the {WiOpt'11} Conf.}, May 2011, pp.
  94--101.

\bibitem{cheng_resource_allocation}
W.~Cheng, X.~Zhang, and H.~Zhang, ``Full/half duplex based resource allocations
  for statistical quality of service provisioning in wireless relay networks,''
  in \emph{Proc. of the {IEEE INFOCOM'12} Conf.}, Mar. 2012, pp. 864--872.

\bibitem{fang_routing_FD}
X.~Fang, D.~Yang, and G.~Xue, ``Distributed algorithms for multipath routing in
  full-duplex wireless networks,'' in \emph{Proc. of the {IEEE MASS'11} Conf.},
  Oct. 2011, pp. 102--111.

\bibitem{FCC_2002}
``Spectrum policy task force,'' \emph{Federal Communications Commission, Rep.
  ET Docket No. 02-135}, Nov. 2002.

\bibitem{Zhao_07_survey}
Q.~Zhao and B.~Sadler, ``A survey of dynamic spectrum access: {S}ignal
  processing, networking, and regulatory policy,'' \emph{{IEEE} Signal
  Processing Magazine}, vol.~24, no.~3, pp. 79--89, May 2007.

\bibitem{haykin}
S.~Haykin, ``Cognitive radio: Brain-empowered wireless communications,''
  \emph{IEEE J. Select. Areas in Commun.}, vol.~23, no.~2, pp. 201--220, Feb.
  2005.

\bibitem{cheng_FD_sensing}
W.~Cheng, X.~Zhang, and H.~Zhang, ``Full duplex spectrum sensing in
  non-time-slotted cognitive radio networks,'' in \emph{Proc. of the
  {MILCOM'11} Conf.}, Nov. 2011, pp. 1029--1034.

\bibitem{cheng_imperfect_fullduplex}
------, ``Imperfect full duplex spectrum sensing in cognitive radio networks,''
  in \emph{Proc. of {CoRoNet'11} Workshop}, Sep. 2011, pp. 1--6.

\bibitem{yuan_mobihoc_07}
Y.~Yuan, P.~Bahl, R.~Chandra, T.~Moscibroda, and Y.~Wu, ``Allocating dynamic
  time-spectrum blocks in cognitive radio networks,'' in \emph{Proc. of the
  {ACM MobiHoc'07} Conf.}, Sep. 2007, pp. 130--139.

\bibitem{vu_scaling_laws}
M.~Vu, N.~Devroye, M.~Sharif, and V.~Tarokh, ``Scaling laws of cognitive
  networks,'' in \emph{Proc. of the {CrownCom'07} Conf.}, Aug. 2007, pp. 2 --8.

\bibitem{sensing_throughput_tradeoff}
Y.-C. Liang, Y.~Zeng, E.~C.~Y. Peh, and A.~T. Hoang, ``Sensing-throughput
  tradeoff for cognitive radio networks,'' \emph{IEEE Transactions on Wireless
  Communications}, vol.~7, no.~4, pp. 1326--1337, April 2008.

\bibitem{Zhao_Quickest_Change_Detection}
J.~Ye and Q.~Zhao, ``Quickest change detection in multiple on-off processes:
  Switching with memory,'' in \emph{Proc. of the {Allerton'09} Conf.}, Oct.
  2009, pp. 1476--1481.

\bibitem{zhao_3}
Q.~Zhao and J.~Ye, ``Quickest detection in multiple on-off processes,''
  \emph{IEEE Transactions on Signal Processing}, vol.~58, no.~12, pp.
  5994--6006, Dec. 2010.

\end{thebibliography}

\end{document}